\documentclass{article}

\usepackage[final,nonatbib]{neurips_2020}

\usepackage[utf8]{inputenc} 
\usepackage[T1]{fontenc}    
\usepackage{hyperref}       
\usepackage{url}            
\usepackage{booktabs}       
\usepackage{amsfonts}       
\usepackage{nicefrac}       
\usepackage{microtype}      
\usepackage{lipsum}
\usepackage{graphicx}
\graphicspath{ {./figures/} }
\usepackage{multirow}
\usepackage{tabularx}

\title{A Systematic Comparison of Encrypted Machine Learning Solutions for Image Classification}

\author{
Veneta Haralampieva\\
  Imperial College London \\
\texttt{veneta.haralampieva19@imperial.ac.uk}
\And
Daniel Rueckert\\
  Imperial College London \\
  Technical University of Munich \\
\texttt{d.rueckert@imperial.ac.uk}\\
\And
Jonathan Passerat-Palmbach \\
  Imperial College London \\
  ConsenSys Health \\
\texttt{j.passerat-palmbach@imperial.ac.uk}
}

\begin{document}
\maketitle
\begin{abstract}
  This work provides a comprehensive review of existing frameworks based on secure computing techniques in the context of private image classification. The in-depth analysis of these approaches is followed by careful examination of their performance costs, in particular runtime and communication overhead.

To further illustrate the practical considerations when using different privacy-preserving technologies, experiments were conducted using four state-of-the-art libraries implementing secure computing at the heart of the data science stack: PySyft and CrypTen supporting private inference via Secure Multi-Party Computation, TF-Trusted utilising Trusted Execution Environments and HE- Transformer relying on Homomorphic encryption.

Our work aims to evaluate the suitability of these frameworks from a usability, runtime requirements and accuracy point of view. In order to better understand the gap between state-of-the-art protocols and what is currently available in practice for a data scientist, we designed three neural network architecture to obtain secure predictions via each of the four aforementioned frameworks. Two networks were evaluated on the MNIST dataset and one on the Malaria Cell image dataset. We observed satisfying performances for TF-Trusted and CrypTen and noted that all frameworks perfectly preserved the accuracy of the corresponding plaintext model.
\end{abstract}


\section{Introduction}
In recent years the use of Deep Learning models has become much more prevalent, with these models achieving outstanding results in a variety of fields such as Computer Vision \cite{redmon2016you} and Natural Language Processing \cite{devlin2018bert}. At the same time, Machine Learning-as-a-service (MLaaS) solutions hosted on cloud platforms are a common strategy for deploying Deep Learning models. They however rely on clients submitting raw data to the server which is unacceptable when utilising sensitive data such as patient records or financial statements.

In certain scenarios, such as medical applications, there are strict rules and regulations governing the storage and usage of personal data such as General Data Protection Regulation (GDPR) in the European Union and Health Insurance Portability and Accountability Act (HIPAA) in the United States. Similarly, the model owners consider the trained model and its parameters to be their own intellectual property and do not want to reveal it to customers or collaborators for inference.

These mutual privacy constraints have prompted the emergence of a body of work leveraging Privacy-Preserving (PP) methods such as Homomorphic Encryption, Secure Multi-Party Computation and Trusted Execution Environments in the context of PP inference. The results exposed in the literature and summarised in Section \ref{sec:related} encouraged us to produce a more practical analysis of the current possibilities of secure inference for image classification. This work fills the following gaps in existing studies: 1) the homogeneity of tested model architectures and datasets, 2) the evaluation of these secure inference techniques in the more realistic context of the most popular deep learning and data science software stack: Python, PyTorch and TensorFlow.

Our study consists in experiments run over two datasets: \textbf{MNIST} \cite{lecun2010mnist}: A standard Machine Learning dataset consisting of 60,000 training and 10,000 testing greyscale images of handwritten digits from 0-9 of size 28x28, and the \textbf{Malaria Cell Images} \cite{malaria2020}: A balanced medical imaging dataset containing 27,558 images with 2 classes: infected and uninfected individuals. The data was shuffled and split randomly into $90\%$ training data and $10\%$ testing data as done previously \cite{riazi2019xonn}.

Section \ref{sec:frameworks} introduces the four Python libraries that were used to compare Private Inference solutions based on Secure Multi-Party Computation, Homomorphic encryption and Trusted Execution Environments. All experiments described in Section \ref{sec:experiments} were conducted by first training a plaintext model and then performing secure inference.

\section{Privacy-preserving inference in the literature}
\label{sec:related}

\subsection{Homomorphic Encryption}
 One of the key seminal works in private inference, proposed by \cite{gilad2016cryptonets} was CryptoNets which relied on Leveled Homomorphic Encryption (LHE) and supported both fully connected and convolutional layers since they can be performed using matrix multiplications. One characteristic disadvantage of the framework was its inability to support commonly used non-linear activations, as well as Max Pooling layers, due to the fact they cannot be computed using polynomials. The authors instead proposed replacing these with HE friendly variants such as using the square activation and a variation of average pooling instead of max pooling. Later, \cite{chao2019carenets} extended their work in CaRENets, where they proposed a new image packing scheme to more efficiently store and process the encrypted data. A radically different approach was explored by \cite{vizitiu2020applying} who suggested using the Matrix Operation for Randomization or Encryption (MORE) scheme \cite{kipnis2012efficient}. Unlike the ring based HE schemes used in previous solutions \cite{gilad2016cryptonets, chao2019carenets} which represented the data as polynomials and relied on noise additions, the MORE algorithm encodes data as a matrix in a noise-free non-deterministic manner. While this approach offered weaker security, it provided an innovative way of handling non-linear activations using the properties of the encrypted matrix. 
 
\subsection{Secure Multi-Party Computation}
Numerous frameworks based on Secure Multi-Party Computation (SMPC) techniques such as Garbled Circuits (GC) \cite{yao1986generate}, Secret Sharing (SS) \cite{shamir1979share} and Oblivious Transfer (OT) \cite{even1985randomized} have been proposed in the context on private inference. All operate in a two party honest-but-curious setting where one party is considered the data owner and the other the model owner. 

SecureML \cite{mohassel2017secureml} was the first framework to provide both secure training and inference. It used SS and Beaver triples \cite{beaver1991efficient}, generated during an offline phase, for matrix multiplications in the fully connected layers and represented non-linear activations as GC, with the sigmoid and softmax activations being approximated.  
Next, \cite{liu2017oblivious} proposed MiniONN, a private inference framework based on SS and GC. Unlike SecureML \cite{mohassel2017secureml}, which implemented only fully connected layers, MiniONN provided support for convolutional ones and implemented them in a privacy-preserving manner using SS. Additive SS was further used for average pooling layers, while max pooling and non-linear activations were done by GC. Similarly, the Chameleon \cite{riazi2018chameleon} framework also relied on Additive SS for matrix multiplications, and used either GC or Goldreich-Micali-Wigderson (GMW) circuits \cite{goldreich1987}, depending on their efficiency, for performing non-linear activations and max pooling. The Gazelle framework \cite{juvekar2018gazelle} instead used an efficient HE scheme for matrix multiplications, and switched to GC for activation functions and pooling layers. 

Further work by \cite{rouhani2018deepsecure} explored relying only on GC for the computation of ReLU activations, fully connected and convolutional layers in the DeepSecure framework. The authors also suggested applying data pre-processing, which projected the data to a low dimensional space and used both during training and inference, and network post-processing, which removed neurons whose activations were below a certain threshold and did not contribute significantly to the accuracy \cite{rouhani2018deepsecure}. Another approach based on GC was the XONN framework \cite{riazi2019xonn}, which was able to achieve a constant communication cost with respect to the depth of the network, making it several orders of magnitude more efficient than prior work \cite{gilad2016cryptonets, liu2017oblivious, rouhani2018deepsecure}. Since it operated on Binarized Neural Networks (BNNs) \cite{courbariaux2016binarized}, where the weights and activations take only binary values (either -1 or 1) \cite{riazi2019xonn}, the costly multiplication operations were replaced with XNOR gates, which are considered free in GC, and Max Pooling layers efficiently computed using OR gates. Similarly to DeepSecure \cite{rouhani2018deepsecure}, pre and post-processing phases could be done to further improve the runtime. 
The SecureNN \cite{wagh2018securenn} and the recent Falcon \cite{wagh2020falcon} frameworks relied exclusively on efficient SS protocols for performing not only matrix multiplications, but also for computing ReLU activations, average and max pooling layers, thus reducing the runtime and communication cost of private training and inference considerably. 

\subsection{Trusted Execution Environments}
Trusted Execution Environments (TEEs) such as Intel's SGX offer a promising avenue for performing private inference due to their isolation from the main operating system, one can securely perform operations inside them. They could be used to store an already pre-trained model which is never visible to the main system and perform private inference. While they offer considerable speed improvements in comparison to HE and SMPC approaches, they have limited memory available and are still considerably slower than standard untrusted CPUs and GPUs. To address these limitations, \cite{tramer2018slalom} proposed Slalom, a framework which efficiently utilises TEEs, where non-linear activations and pooling operations are executed, and a co-located untrusted hardware which can be used to securely perform the costly matrix computation of fully connected and convolutional layers. 

\subsection{Comparison of different approaches}
Most of the works from the literature provide performance benchmarks against four main neural networks evaluated on the MNIST dataset. 
Both \textbf{Network A} \cite{mohassel2017secureml} and \textbf{Network B} \cite{gilad2016cryptonets} used the square activation, however, the former included 3 fully connected layers, while the latter had a Convolutional and two linear layers. 
\textbf{Network C} consisted of a single convolutional and two fully connected layers with ReLUs. Finally, \textbf{Network D} had two sets of convolutions and max pooling layers, followed by two linear layers and used ReLU throughout. It should be noted that the XONN framework \cite{riazi2019xonn} had only binary activations. 

\begin{table*}[ht!]
\centering
\begin{tabular}{l l l l lll lll}
\toprule
  \multirow{2}{*}{\textbf{Framework}} & \multirow{2}{*}{\textbf{Model}} & \multirow{2}{*}{\textbf{Parties}} & \multirow{2}{*}{\textbf{Accuracy}} &  \multicolumn{3}{c}{\textbf{Runtime (s)}} & \multicolumn{3}{c}{\textbf{Comm (MB)}}  \\
  
  & & & & Offline & Online & Total & Offline & Online & Total  \\
  \midrule
  SecureML & \multirow{6}{*}{A} & 2PC &  $93.1\%$ & 4.7 & 0.18 & 4.88 & - & - & - \\
  MiniONN & & 2PC & $97.6\%$ & 0.9 & 0.14 & 1.04 & 3.8 & 12 & 15.8 \\
  Gazelle &  & 2PC & $97.6\%$ & - & 0.03 & 0.03 & - & 0.5 & 0.5 \\
  SecureNN & & 3PC & $93.4\%$ & - & 0.043 & 0.043 & - & 2.1 & 2.1 \\
  Falcon & & 3PC & $93.4\%$ & \textbf{-} & \textbf{0.011} & \textbf{0.011} & \textbf{-} & \textbf{0.012} & \textbf{0.012} \\
  XONN & & 2PC & $97.6\%$ & - & 0.13 & 0.13 & - & 4.29 & 4.29 \\
  \midrule
  CryptoNets & \multirow{3}{*}{B} & 2PC & $98.95\%$ & - & 297.5 & 297.5 & - & 372.2 & 372.2 \\
  MiniONN & & 2PC & $98.95\%$ & 0.88 & 0.4 & 1.28 & 3.6 & 44 & 47.6 \\
  Gazelle &  & 2PC & $97.6\%$ & \textbf{-} & \textbf{0.03} & \textbf{0.03} & \textbf{-} & \textbf{0.5} & \textbf{0.5} \\
  XONN & & 2PC & $98.64\%$ & - & 0.16 & 0.16 & - & 38.28 & 38.28 \\
  \midrule
  DeepSecure & \multirow{6}{*}{C} & 2PC & $98.95\%$ & - & 9.67 & 9.67 & - & 791 & 791 \\
  Chameleon & & 3PC & $99.0\%$ & 1.25 & 0.99 & 2.24 & 5.4 & 5.1 & 10.5 \\
  Gazelle &  & 2PC & $99.0\%$ & 0.15 & 0.05 & 0.20 & 5.9 & 2.1 & 8.0 \\
  SecureNN &  & 3PC & $99.0\%$ & - & 0.076 & 0.076 & - & 4.05 & 4.05 \\
  Falcon & & 3PC & $98.77\%$ & \textbf{-} & \textbf{0.009} & \textbf{0.009} & \textbf{-} & \textbf{0.049} & \textbf{0.049} \\
  XONN & & 2PC & $98.64\%$ & - & 0.16 & 0.16 & - & 38.28 & 38.28 \\
  \midrule
  MiniONN & \multirow{5}{*}{D} & 2PC & $99.31\%$ & 3.58 & 5.74 & 9.32 & 20.9 & 636.6 & 657.5 \\
  Gazelle & & 2PC & $99.0\%$ & 0.481 & 0.33 & 0.811 & 47.5 & 22.5 & 70.0 \\
  SecureNN & & 3PC & $99.15\%$ & - & 0.13 & 0.13 & - & 8.86 & 8.86 \\
  Falcon & & 3PC & $99.15\%$ & \textbf{-} & \textbf{0.042} & \textbf{0.042} & \textbf{-} & \textbf{0.51} & \textbf{0.51} \\
  XONN & & 2PC & $99.0\%$ & - & 0.15 & 0.15 & - & 32.13 & 32.13 \\
  \bottomrule
  \end{tabular}
\caption[\textbf{Performance benchmarks obtained on the MNIST dataset of different privacy-preserving frameworks when executing private inference on a single image (LAN setting).}]{\textbf{Performance benchmarks obtained on the MNIST dataset of different privacy-preserving frameworks when executing private inference on a single image (LAN setting).}}
\label{tablesummary}
\end{table*}

Table \ref{tablesummary} demonstrates that the research in privacy-preserving frameworks has lead to considerable reduction in communication and improvements in performance. When examining a deeper model like Network D, Falcon \cite{wagh2020falcon} is over 200X faster than MiniONN  \cite{liu2017oblivious} and requires over 1000X less memory. While solutions based entirely on HE generally perform worse, a mixed approach like Gazelle \cite{juvekar2018gazelle} showcases an efficient combination of HE and SMPC. The difference in the accuracy reported by different frameworks can be attributed to varying hyper-parameters used during training and the application of specific techniques, such as batch normalisation in Network A by MiniONN \cite{liu2017oblivious}.

Several frameworks did not provide benchmarks against these existing architectures and are not present in Table \ref{tablesummary}, however, they discussed improvements with respect to other baselines. CaRENets \cite{chao2019carenets} reported a 5x memory usage improvement on the MNIST dataset in comparison to CryptoNets, while the Slalom framework \cite{tramer2018slalom} achieved significant performance improvements on larger models such as VGG16 \cite{simonyan2014very} and ResNet \cite{he2016deep} when compared to a solution where all computations are performed inside the secure enclave \cite{tramer2018slalom}. The approach proposed by \cite{vizitiu2019towards}, based on the MORE scheme, was successfully applied to several deep models in the medical imaging domain.



\section{Privacy-Preserving Inference in practice}

\subsection{Frameworks}
\label{sec:frameworks}

\subsubsection{PySyft \cite{ryffel2018generic}}
A prominent private learning library which implements the SPDZ \cite{damgaard2012multiparty} and SecureNN protocols \cite{wagh2018securenn} for private inference. It supports two-party SMPC and relies on a trusted third party, the crypto provider, for generating correlated randomness using the Beaver triplets technique \cite{beaver1991efficient}. As discussed by \cite{wagh2018securenn}, floating point numbers cannot be used directly, which is why PySyft uses fixed precision encoding \cite{ryffel2018generic}. Each party is represented as either a virtual or a network worker. The former is instantiated on the same device and simulates network communications, while the latter can be run on the same device or different ones and communicates over Web Sockets \cite{ryffel2018generic}. Both worker types apply the same serialization and deserialization steps. In this work, Virtual Workers were used for simplicity as they still allow the number of communication bytes sent between parties to be measured.

\subsubsection{CrypTen \cite{crypten2019}} A recent private inference library with support for PyTorch which can be used for both Private Training and Inference. While there is no official white paper published yet, the presentation given at NeurIPS 2019 outlines that CrypTen utilises both secret sharing (for matrix multiplications) and garbled circuits (for evaluating non-linearities) and includes protocols for converting between these two secret sharing schemes \cite{cryptenslides2019}. The communication between parties is implemented using the Gloo backend, a PyTorch distributed backend where each party runs the same code in a different process. Similar to PySyft it also uses a trusted third party for computing Beaver triplets and operates in an honest-but-curious setting \cite{cryptenslides2019}. 

\subsubsection{TF-Trusted \cite{tftrusted2019}}
A Tensorflow framework which allows models to be executed within an Intel SGX secure enclave and is based on the Asylo library, a generic framework for executing computations on secure enclaves \cite{asylo2018}. A pre-trained Tensorlow model, saved using the protobuf format, can be used for private inference. For the purpose of this benchmark, we have added support for Tensorflow 1.14.0 in place of 1.13.1 to compare identical TF versions.

\subsubsection{HE-Transformer}
Also known as nGraph-HE2 \cite{boemer2019ngraph}, it is a library for private inference using Homomorphic Encryption (HE) built using Tensorflow 1.14.0, Intel nGraph's Deep Learning model compiler \cite{cyphers2018intel} and Microsoft SEAL version 3.3.0 which implements the CKKS scheme \cite{sealcrypto}. It operates in a two party setting: a server who is the model owner and the client who holds data. Additionally, the client also performs non-linear activations such as ReLU and Max Pooling layers. After computing a linear or a convolutional layer the server sends the encrypted data to the client, who decrypts it, applies the function, re-encrypts it and returns the result \cite{boemer2019ngraph}. Similarly to Gazelle \cite{juvekar2018gazelle}, that process helps manage the noise growth by refreshing the ciphertext after every layer. 

\subsection{Conducted experiments}
\label{sec:experiments}


\subsubsection{Model architectures}
Three main architectures were evaluated, inspired by the ones reviewed in section \ref{sec:related}. \textbf{Model X and Y} were evaluated on the MNIST dataset, while \textbf{Model Z} was applied to the Malaria dataset which is more complex and required a deeper model. Model X consists of three fully connected layers, while Model Y inserts both a convolution and a pooling operation before two fully connected ones. Finally, Model Z, based on the network proposed by \cite{riazi2019xonn}, includes two sets of convolution and pooling layers followed by two fully connected ones.

\subsubsection{Experimental setting}
Both datasets were normalised to have zero mean and unit standard deviation prior to using them for training or inference. The Malaria dataset was resized to 32x32 similar to \cite{riazi2019xonn} which enables faster inference time. All experiments were performed on a machine with a 2.6 GHz 8-Core Intel Xeon E5-2650 v2 and 256 GB RAM. The \textit{tf-trusted} experiments that require an Intel SGX chip were run on a 2.80GHz Intel Core i7-7600U CPU and 16GB RAM. Each runtime benchmark was executed 20 times and the average time was reported in seconds. All three networks were optimised using Adam \cite{kingma2014adam} with a learning rate of $0.001$. The MNIST models were trained with a batch size of $128$ for $15$ epochs, while the Malaria one using batch size $52$ and $20$ epochs. The associated code is available here \footnote{\url{https://github.com/venetahar/iso-privacy-smpc}}\footnote{\url{https://github.com/venetahar/iso-privacy-tee-he}}. 

\subsubsection{Results}
\begin{table}[ht]
    \centering
  \begin{tabular}{ccccl}
    \toprule
    Library  & Model & Accuracy & Runtime (s) & Comm (MB)\\
    \midrule
    PyTorch  & \multirow{6}{*}{X} & $97.66\%$ & 0.0002 & -\\
    PySyft & & $97.66\%$ & 0.5 & 5.71\\
    CrypTen & & $97.66\%$ & 0.043 & 1.01\\
    Tensorflow & & $97.63\%$ & 0.0003 & -\\
    TF-Trusted & & $97.63\%$ & 0.14 & -\\
    HE-Transformer & & $97.63\%$ & 15.4 & 1171.94\\
  \midrule
  PyTorch & \multirow{6}{*}{Y} & $98.47\%$ & 0.0003 & -\\
  PySyft & & $98.47\%$ & 0.66 & 1.28\\
  CrypTen & & $98.47\%$ & 0.035 & 0.35\\
  Tensorflow & & $98.05\%$ & 0.0003 & -\\
  TF-Trusted & & $98.05\%$ & 0.12 & -\\
  HE-Transformer & & $98.05\%$ & 12.6 & 2780.66\\
  \midrule
    PyTorch & \multirow{6}{*}{Z} & $94.85\%$ & 0.0011 & -\\
  PySyft  & & $94.85\%$ & 8.26 & 38.56\\
  CrypTen & & $94.85\%$ & 0.25 & 8.29\\
  Tensorflow & & $91.5\%$ & 0.0008 & -\\
  TF-Trusted & & $91.5\%$ & 0.14 & -\\
  HE-Transformer & & $91.5\%$ & 548.94 & 56703.68\\
  \bottomrule
\end{tabular}
\caption{\textbf{Performance benchmarks during inference using a single data instance on MNIST \cite{lecun2010mnist} (Models X and Y) and the Malaria dataset \cite{malaria2020} (Model Z).}}
  \label{tab:experimentsmnist}
\end{table}



As it can be observed in Table \ref{tab:experimentsmnist} \textbf{all frameworks preserve the accuracy} of the plain text model when performing private inference. It can be seen that CrypTen is significantly faster than PySyft which is clearer  in Model Z. This is an interesting find as CrypTen requires share conversions between Additive SS and GCs \cite{cryptenslides2019} which has been previously reported as inefficient \cite{wagh2018securenn}. We attribute the discrepancy in runtimes to the use of different communication backends and the serialization and deserialization applied. Additionally, the results show that CrypTen requires less bandwidth which could be explained by the use of Garbled Circuits to represent ReLU activations while the PySyft implementation of ReLU requires more communication between parties for the additive shares to be exchanged. While there are inherent performance differences between Tensorflow and PyTorch, it can be seen that TF-Trusted is able to evaluate larger models such as Model Z much faster. Intuitively this makes sense as the inference is performed inside the enclave using a plain text version of the data. While HE-Transformer also preserves the accuracy, this comes at a much higher computational and memory cost mainly due to the communication with the client after every fully connected or convolutional layer where the non-linearity is applied.



\section{Conclusion}

Overall, the experiments conducted demonstrate that it is possible to leverage PPML methods purely from the traditional data science stack. Our experiments have produced a unified codebase only diverging at the underlying deep-learning framework level: PyTorch or TensorFlow, and could go even further by manipulating ONNX models in the future. Most importantly, model accuracy was preserved compared to the plaintext version, regardless of the PPML method employed, demonstrating that the studied frameworks abstract the low-level implementation constraints, such as fixed-precision arithmetic, in a satisfying manner.

With regards to run times, Crypten is worth noting for displaying performances very close to the hardware-based solution of TF-Trusted and as such might appeal to a broader range of users by lifting the hardware requirement and memory space constraints imposed by Intel SGX. On the other hand, HE-Transformer suffers from the genericity of its approach, since a bespoke protocol would have fared better than the automatic toolchain. Communication and memory requirements currently make the HE option impractical. As part of our future iterations on this work, we will thus integrate other secure inference scenarios evaluating batched predictions, without real-time constraints. This setting and further manual optimisations might prove a better fit for HE-based solutions. It will also be interesting to evaluate how other optimisations on the model side, such as quantisation, impact the results for each framework.

\vfill
\newpage


\begin{thebibliography}{10}

\bibitem{redmon2016you}
Joseph Redmon, Santosh Divvala, Ross Girshick, and Ali Farhadi.
\newblock You only look once: Unified, real-time object detection.
\newblock In {\em Proceedings of the IEEE conference on computer vision and
  pattern recognition}, pages 779--788, 2016.

\bibitem{devlin2018bert}
Jacob Devlin, Ming-Wei Chang, Kenton Lee, and Kristina Toutanova.
\newblock Bert: Pre-training of deep bidirectional transformers for language
  understanding.
\newblock {\em arXiv preprint arXiv:1810.04805}, 2018.

\bibitem{lecun2010mnist}
Yann LeCun, Corinna Cortes, and CJ~Burges.
\newblock Mnist handwritten digit database.
\newblock {\em ATT Labs [Online]. Available: http://yann. lecun.
  com/exdb/mnist}, 2, 2010.

\bibitem{malaria2020}
Sivaramakrishnan Rajaraman, Sameer~K Antani, Mahdieh Poostchi, Kamolrat
  Silamut, Md~A Hossain, Richard~J Maude, Stefan Jaeger, and George~R Thoma.
\newblock Pre-trained convolutional neural networks as feature extractors
  toward improved malaria parasite detection in thin blood smear images.
\newblock {\em PeerJ}, 6:e4568, 2018.

\bibitem{riazi2019xonn}
M~Sadegh Riazi, Mohammad Samragh, Hao Chen, Kim Laine, Kristin Lauter, and
  Farinaz Koushanfar.
\newblock Xonn: Xnor-based oblivious deep neural network inference.
\newblock In {\em 28th USENIX Security Symposium (USENIX Security 19)}, pages
  1501--1518, 2019.

\bibitem{gilad2016cryptonets}
Ran Gilad-Bachrach, Nathan Dowlin, Kim Laine, Kristin Lauter, Michael Naehrig,
  and John Wernsing.
\newblock Cryptonets: Applying neural networks to encrypted data with high
  throughput and accuracy.
\newblock In {\em International Conference on Machine Learning}, pages
  201--210, 2016.

\bibitem{chao2019carenets}
Jin Chao, Ahmad~Al Badawi, Balagopal Unnikrishnan, Jie Lin, Chan~Fook Mun,
  James~M Brown, J~Peter Campbell, Michael Chiang, Jayashree Kalpathy-Cramer,
  Vijay~Ramaseshan Chandrasekhar, et~al.
\newblock Carenets: compact and resource-efficient cnn for homomorphic
  inference on encrypted medical images.
\newblock {\em arXiv preprint arXiv:1901.10074}, 2019.

\bibitem{vizitiu2020applying}
Anamaria Vizitiu, Cosmin~Ioan Nitǎ, Andrei Puiu, Constantin Suciu, and
  Lucian~Mihai Itu.
\newblock Applying deep neural networks over homomorphic encrypted medical
  data.
\newblock {\em Computational and Mathematical Methods in Medicine}, 2020, 2020.

\bibitem{kipnis2012efficient}
Aviad Kipnis and Eliphaz Hibshoosh.
\newblock Efficient methods for practical fully homomorphic symmetric-key
  encrypton, randomization and verification.
\newblock {\em IACR Cryptology ePrint Archive}, 2012:637, 2012.

\bibitem{yao1986generate}
Andrew Chi-Chih Yao.
\newblock How to generate and exchange secrets.
\newblock In {\em 27th Annual Symposium on Foundations of Computer Science
  (sfcs 1986)}, pages 162--167. IEEE, 1986.

\bibitem{shamir1979share}
Adi Shamir.
\newblock How to share a secret.
\newblock {\em Communications of the ACM}, 22(11):612--613, 1979.

\bibitem{even1985randomized}
Shimon Even, Oded Goldreich, and Abraham Lempel.
\newblock A randomized protocol for signing contracts.
\newblock {\em Communications of the ACM}, 28(6):637--647, 1985.

\bibitem{mohassel2017secureml}
Payman Mohassel and Yupeng Zhang.
\newblock Secureml: A system for scalable privacy-preserving machine learning.
\newblock In {\em 2017 IEEE Symposium on Security and Privacy (SP)}, pages
  19--38. IEEE, 2017.

\bibitem{beaver1991efficient}
Donald Beaver.
\newblock Efficient multiparty protocols using circuit randomization.
\newblock In {\em Annual International Cryptology Conference}, pages 420--432.
  Springer, 1991.

\bibitem{liu2017oblivious}
Jian Liu, Mika Juuti, Yao Lu, and Nadarajah Asokan.
\newblock Oblivious neural network predictions via minionn transformations.
\newblock In {\em Proceedings of the 2017 ACM SIGSAC Conference on Computer and
  Communications Security}, pages 619--631, 2017.

\bibitem{riazi2018chameleon}
M~Sadegh Riazi, Christian Weinert, Oleksandr Tkachenko, Ebrahim~M Songhori,
  Thomas Schneider, and Farinaz Koushanfar.
\newblock Chameleon: A hybrid secure computation framework for machine learning
  applications.
\newblock In {\em Proceedings of the 2018 on Asia Conference on Computer and
  Communications Security}, pages 707--721, 2018.

\bibitem{goldreich1987}
Oded Goldreich, S.~Micali, and Avi Wigderson.
\newblock How to play any mental game.
\newblock pages 218--229, 01 1987.

\bibitem{juvekar2018gazelle}
Chiraag Juvekar, Vinod Vaikuntanathan, and Anantha Chandrakasan.
\newblock Gazelle: A low latency framework for secure neural network inference.
\newblock In {\em 27th USENIX Security Symposium (USENIX Security 18)}, pages
  1651--1669, 2018.

\bibitem{rouhani2018deepsecure}
Bita~Darvish Rouhani, M~Sadegh Riazi, and Farinaz Koushanfar.
\newblock Deepsecure: Scalable provably-secure deep learning.
\newblock In {\em Proceedings of the 55th Annual Design Automation Conference},
  pages 1--6, 2018.

\bibitem{courbariaux2016binarized}
Matthieu Courbariaux, Itay Hubara, Daniel Soudry, Ran El-Yaniv, and Yoshua
  Bengio.
\newblock Binarized neural networks: Training deep neural networks with weights
  and activations constrained to+ 1 or-1.
\newblock {\em arXiv preprint arXiv:1602.02830}, 2016.

\bibitem{wagh2018securenn}
Sameer Wagh, Divya Gupta, and Nishanth Chandran.
\newblock Securenn: Efficient and private neural network training.
\newblock {\em IACR Cryptology ePrint Archive}, 2018:442, 2018.

\bibitem{wagh2020falcon}
Sameer Wagh, Shruti Tople, Fabrice Benhamouda, Eyal Kushilevitz, Prateek
  Mittal, and Tal Rabin.
\newblock Falcon: Honest-majority maliciously secure framework for private deep
  learning.
\newblock {\em arXiv preprint arXiv:2004.02229}, 2020.

\bibitem{tramer2018slalom}
Florian Tramer and Dan Boneh.
\newblock Slalom: Fast, verifiable and private execution of neural networks in
  trusted hardware.
\newblock {\em arXiv preprint arXiv:1806.03287}, 2018.

\bibitem{simonyan2014very}
Karen Simonyan and Andrew Zisserman.
\newblock Very deep convolutional networks for large-scale image recognition.
\newblock {\em arXiv preprint arXiv:1409.1556}, 2014.

\bibitem{he2016deep}
Kaiming He, Xiangyu Zhang, Shaoqing Ren, and Jian Sun.
\newblock Deep residual learning for image recognition.
\newblock In {\em Proceedings of the IEEE conference on computer vision and
  pattern recognition}, pages 770--778, 2016.

\bibitem{vizitiu2019towards}
Anamaria Vizitiu, Cosmin~Ioan Nitǎ, Andrei Puiu, Constantin Suciu, and
  Lucian~Mihai Itu.
\newblock Towards privacy-preserving deep learning based medical imaging
  applications.
\newblock In {\em 2019 IEEE International Symposium on Medical Measurements and
  Applications (MeMeA)}, pages 1--6. IEEE, 2019.

\bibitem{ryffel2018generic}
Theo Ryffel, Andrew Trask, Morten Dahl, Bobby Wagner, Jason Mancuso, Daniel
  Rueckert, and Jonathan Passerat-Palmbach.
\newblock A generic framework for privacy preserving deep learning.
\newblock {\em arXiv preprint arXiv:1811.04017}, 2018.

\bibitem{damgaard2012multiparty}
Ivan Damg{\aa}rd, Valerio Pastro, Nigel Smart, and Sarah Zakarias.
\newblock Multiparty computation from somewhat homomorphic encryption.
\newblock In {\em Annual Cryptology Conference}, pages 643--662. Springer,
  2012.

\bibitem{crypten2019}
Facebook AI~Research Revision.
\newblock Crypten.
\newblock \url{https://crypten.readthedocs.io/en/latest/}, 2019.
\newblock Accessed: 2020-03-28.

\bibitem{cryptenslides2019}
Shubho Sengupta.
\newblock Private ai.
\newblock \url{https://research.fb.com/neurips-2019-expo-workshops/}, 2019.
\newblock Accessed: 2020-03-28.

\bibitem{tftrusted2019}
Cape Privacy (Formerly~Dropout Labs).
\newblock Tf trusted.
\newblock \url{https://github.com/capeprivacy/tf-trusted}, 2019.
\newblock Accessed: 2020-03-28.

\bibitem{asylo2018}
Google).
\newblock Asylo.
\newblock \url{https://asylo.dev}, 2018.
\newblock Accessed: 2020-03-28.

\bibitem{boemer2019ngraph}
Fabian Boemer, Anamaria Costache, Rosario Cammarota, and Casimir Wierzynski.
\newblock ngraph-he2: A high-throughput framework for neural network inference
  on encrypted data.
\newblock In {\em Proceedings of the 7th ACM Workshop on Encrypted Computing \&
  Applied Homomorphic Cryptography}, pages 45--56, 2019.

\bibitem{cyphers2018intel}
Scott Cyphers, Arjun~K Bansal, Anahita Bhiwandiwalla, Jayaram Bobba, Matthew
  Brookhart, Avijit Chakraborty, Will Constable, Christian Convey, Leona Cook,
  Omar Kanawi, et~al.
\newblock Intel ngraph: An intermediate representation, compiler, and executor
  for deep learning.
\newblock {\em arXiv preprint arXiv:1801.08058}, 2018.

\bibitem{sealcrypto}
{M}icrosoft {SEAL} (release 3.3).
\newblock \url{https://github.com/Microsoft/SEAL}, 2019.
\newblock Microsoft Research, Redmond, WA.

\bibitem{kingma2014adam}
Diederik~P. Kingma and Jimmy Ba.
\newblock Adam: A method for stochastic optimization, 2014.

\end{thebibliography}

\end{document}